\documentstyle[psbox]{WORKSHOP}
\begin{document}

\newcommand{\xmm}{{\it XMM}}
\newcommand{\asca}{{\it ASCA}}
\title{
The inverse iron-bias in action in Abell 2028
}

\author{
Fabio Gastaldello$^{1,2,3}$, Stefano Ettori$^{4,5}$, Italo Balestra$^6$,\\
Fabrizio Brighenti$^{7}$, David Buote$^3$, Sabrina De Grandi$^8$, Myriam Gitti$^{4,9}$ and Paolo Tozzi$^{10}$
\\[12pt]  
%
$^1$  INAF, IASF, via Bassini 15, I-20133 Milano, Italy $^2$  Occhialini Fellow \\
$^3$  University of California Irvine, 4129, Frederick Reines Hall, Irvine, CA 92697, USA \\
$^4$  INAF, Osservatorio Astronomico di Bologna, via Ranzani 1, I-40127 Bologna, Italy \\
$^5$  INFN, Sezione di Bologna, viale Berti Pichat 6/2, I-40127 Bologna, Italy \\
$^6$  MPE, Karl-Schwarzschild-Str. 2, D-85741 Garching, Germany \\
$^7$  Dipartimento di Astronomia, Universit\`a di Bologna, via Ranzani 1, Bologna 40127, Italy \\
$^8$  INAF, Osservatorio Astronomico di Brera, via E. Bianchi 46, I-23807 Merate(LC), Italy \\
$^9$  Harvard-Smithsonian Center for Astrophysics, 60 Garden Street, Cambridge, MA 02138, USA \\
$^{10}$  INAF, Osservatorio Astronomico di Trieste, via G.B. Tiepolo 11, I-34131 Trieste, Italy \\
%
{\it E-mail(FG): gasta@lambrate.inaf.it} 
}

\abst{
Recent work based on a global measurement of the ICM properties find evidence for an increase
of the iron abundance in galaxy clusters with temperature around 2-4 keV. We have undertaken a study of 
the metal distribution in nearby clusters in this temperature range, aiming at resolving 
spatially the metal content of the ICM. 
The \xmm\ observation of the first object of the sample, 
the cluster Abell 2028, reveals a complex structure of the cluster over scale of $\sim300$
kpc, showing an interaction between two sub-clusters in a ``cometary'' configuration. 
We show that a naive one-component fit for the core of Abell~2028 returns a biased high 
metallicity. This is due to the inverse iron-bias, which is not related to the 
presence in the spectrum of both Fe-L and Fe-K emission lines but to the behavior of
the fitting code in shaping the Fe-L complex of a one temperature component to adjust to the 
multi-temperature structure of the projected spectrum.
}

\kword{galaxies: cluster: general -- intergalactic medium -- X-ray: galaxies. }

\maketitle
\thispagestyle{empty}

\section{Introduction}

The X-ray determination of elemental abundances in the hot
gas of clusters of galaxies is in principle robust because equivalent widths of the observed 
emission lines can be directly converted into the corresponding elemental abundances 
(e.g., Mushotzky et al. 1996). 
Unfortunately the intrinsic simplicity of the measurement has faced the limitations of the 
X-ray satellites in terms of the shape of the instrumental response, bandpass, spectral and 
spatial resolution. The correct modeling of the temperature structure is crucial, in particular
when dealing with spectra extracted from a large aperture centered on the core of clusters and 
groups, where photons coming from regions of different temperature and abundances are mixed 
together, given the presence of strong opposite gradients (i.e. cooler regions are more metal 
rich) in the temperature and metallicity profiles (e.g., De Grandi \& Molendi 2001).\\
The first important recognition of a bias in the measurement of elemental
abundances was the description of the "iron bias" (Buote 2000). The previously found significant subsolar 
values for the iron abundance in groups and elliptical galaxies stem from fitting a multi 
temperature plasma with a simple single temperature model, resulting in best-fitting elemental 
abundance biased low.\\
Another trend in the ICM abundance versus cluster temperature has been shown with
increasing evidence in recent years. Using the \asca\ archive observations of 273 objects 
Baumgartner et al. (2005) showed that clusters with gas temperature between 2 and 4 keV have a 
typical mean abundance that is larger by up to a factor of 3 than hotter systems. 
Rasia et al. (2008) made a fundamental step in recognizing a possible bias analyzing 
mock \xmm\ observations of simulated galaxy clusters and finding a 
systematic overestimate of iron for systems in the 2-3 keV range. They explained it as due to the 
fact that temperatures in the range of 3 keV are very close to the transition between the 
relative importance of the lines (Fe-L or Fe-K) used in determining the global iron content. 
Projection and low resolution effects can create a complex temperature structure, 
averaging in the same region different temperatures with different contributions of Fe-L and Fe-K 
emission. Simionescu et al. (2009) supported this explanation by analyzing a deep \xmm\ observation of 
the high luminosity cluster Hydra~A finding a biased high Fe abundance at a level 
of 35\% in its core, dubbing for the first time this kind of bias the ``inverse'' Fe bias.\\
With the aim of shedding more light on these issues by going beyond single aperture measurements, 
we started with the present work a study of the the metal distribution in iron-richest systems
of Baumgartner et al. (2005), selecting the 9 objects with $Z>0.6 Z_{\odot}$ 
(Anders \& Grevesse 1989 units).
Here we present the analysis of an \xmm\ observation of one of these objects, Abell 2028.

\begin{figure}
\begin{center}
\psbox[xsize=6cm]
{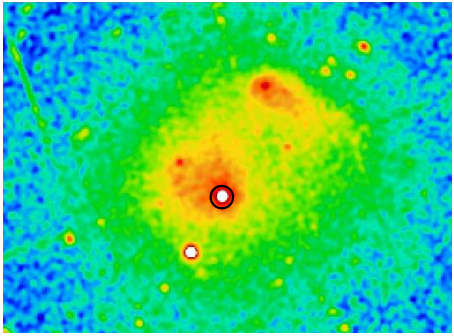} 
\caption{EPIC image of A 2028 in the 0.5-2 keV energy band. The black circle indicate the core region 
used to extract the spectra of Fig.\ref{fig:spectra}, discussed in the text.
} \label{fig:ima}
\end{center}
\end{figure}

\begin{figure}
\begin{center}
\psbox[xsize=8cm]
{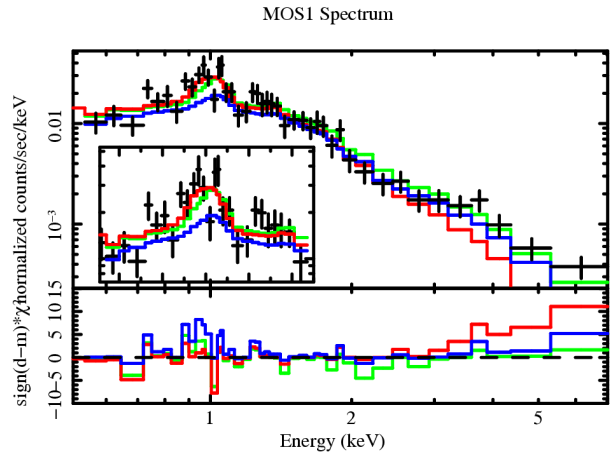} 
\caption{MOS 1 spectrum of the core. With the solid green line the best fit model and residuals of the
1T fit obtained in the broad 0.5-10 keV band are shown; with the red line the fit with 1T model in 
the 0.5-3 keV band; with the blue line the fit with the 1T model with the parameters obtained by the 0.5-10 keV fit
but with the abundance value kept fixed at the value obtained in the 0.5-3 keV band. 
} \label{fig:spectra}
\end{center}
\end{figure}
\vspace{-0.5cm}
\section{The \xmm\ observation of Abell 2028}

The \xmm\ observation of A 2028 clearly
reveals two merging subclusters, a more luminous South-West component and a less luminous
North-East one (see Fig.\ref{fig:ima}). 
The ``cometary'' shape of the two subclusters seems to suggest the direction
of the merging, with the SW main subcluster coming from the NW direction and the smaller
subcluster coming from SE. The temperature map obtained with a single temperature model 
reveals that both subclusters have a cool core with a temperature of 3 keV for the main subcluster and
2 keV for the smaller subcluster and they reach a peak temperature of $\sim5$ keV and $\sim4$ keV 
respectively.

If we focus on the spectrum of the core of the main subcluster, corresponding to the black circle
in Fig.\ref{fig:ima}, which is a circular region of 0.5$^{\prime}$ radius, corresponding to 44 kpc. 
Even in this highest S/N spectrum the Fe-K line complex is not evident but still a single 
temperature fit returns a high metallicity (see Table \ref{tab:spectra}). The 1T model does not
provide a good fit to the data, which clearly indicate the need for a range of temperatures from 
$\sim$2 keV to 4-5 keV, and overestimates the abundance due to the inverse-Fe bias 
(see Tab.\ref{tab:spectra} and Fig.\ref{fig:spectra}).
The inverse-Fe bias is mainly driven by the behavior of the the Fe-L complex strength which
falls rapidly with increasing temperature above roughly 3 keV: in a multi-T spectrum where the average 
T is near 3 keV, but there are temperature components below and above 3 keV, most of the the Fe L 
emission of this multi-T spectrum comes from the lower-T components in the spectrum. However, when one 
fits a (wrong) single-T model, the single temperature will be sufficiently high that
the Fe L lines will be weaker, and thus the model will have to
compensate by increasing the Fe abundance above the true value.

\begin{table}[h]
\caption{Results of the fits using 1T and 2T models in different energy bands for the EPIC spectra of 
the core of A 2028.}
\label{tab:spectra}
\begin{center}
\begin{tabular}{lcccc} \hline\hline\\[-6pt]
Model/Band & $kT_h$ &  Z  &  $kT_c$ & $\chi^2$/dof \\   
\hspace{0.8cm} (keV)        & (keV)    &  ($Z_{\odot}$) & (keV) & \\
\hline
1T 0.5-10    & $3.0\pm0.2$ & $0.76\pm0.15$  &      & 252/168 \\
1T 0.5-3     & $2.0\pm0.2$ & $0.33\pm0.06$  &      & 179/148 \\
2T 0.5-10    & 5.0 (fixed) & $0.46\pm0.07$  & $1.4\pm0.1$    & 206/167 \\
\end{tabular}
\end{center}
\end{table}

\vspace{-0.5cm}
\section*{References}

\re
Anders \& Grevesse 1989, Geo. Cosmo. Acta, 53, 197 

\re
Buote, D.A. 2000, MNRAS, 311, 176

\re
Baumgartner, W.H. et al. 2005, ApJ, 620, 680

\re
De Grandi \& Molendi. 2001, ApJ, 551, 153

\re
Mushotzky, R. et al. 1996, ApJ, 466, 686

\re
Rasia, E. et al. 2008, ApJ, 674, 728

\re
Simionescu, A. et al. 2009, A\&A, 493, 409

\label{last}

\end{document}